\documentclass[reprint,
amsmath,amssymb,
pre,
]{revtex4-2}
\usepackage{CJK}

\usepackage{graphicx}% Include figure files
\usepackage{dcolumn}% Align table columns on decimal point
\usepackage{bm}% bold math
\usepackage{mathtools}
\usepackage{hyperref} % Add hyperref package
\hypersetup{
  colorlinks=true,    % Enables coloring of text instead of using boxes
  linkcolor=blue,     % Color for normal internal links
  citecolor=black,    % Color for bibliographic citations in text
  urlcolor=black       % Color for email addresses, hypertext links, etc.
}
\usepackage{color}
\usepackage{xcolor}
\definecolor{revision_Oct}{RGB}{0, 0, 255}

\def\nn{\nonumber}

\def\el{{\it et al.\ }}

\begin{document}

%\preprint{APS/123-QED}

\begin{CJK*}{UTF8}{gbsn}
\title{Self-consistent expansion and field-theoretic renormalization group for a singular nonlinear diffusion equation with anomalous scaling}
% Force line breaks with \\
\author{Minhui Zhu (朱{\CJKfamily{bsmi}旻}晖)$^1$ and Nigel Goldenfeld$^{1,2}$}
\affiliation{$^1$Department of Physics, University of Illinois at
Urbana-Champaign, Loomis Laboratory of Physics, 1110 West Green
Street, Urbana, Illinois, 61801-3080, USA\\
$^2$Department of Physics, University of California, San Diego, 9500 Gilman Drive, La Jolla, California 92093, USA}

\begin{abstract}
The method of self-consistent expansions is a powerful tool for handling strong coupling problems that might otherwise be beyond the reach of perturbation theory, providing surprisingly accurate approximations even at low order.  First applied in its embryonic form to fully-developed turbulence, it has subsequently been successfully applied to a variety of problems that include polymer statistics, interface dynamics and high order perturbation theory for the anharmonic oscillator.  Here we show that the self-consistent expansion can be applied to singular perturbation problems arising in the theory of partial differential equations in conjunction with renormalization group methods.  We demonstrate its application to Barenblatt's nonlinear diffusion equation for porous media filtration, where the long-time asymptotics exhibits anomalous dimensions that can be systematically calculated using the perturbative renormalization group. We find that even the first-order self-consistent expansion, when combined with the Callan-Symanzik equation, improves the approximation of the anomalous dimension obtained by the first order perturbative renormalization group,  especially in the strong coupling regime. We also develop a field-theoretic framework for deterministic partial differential equations to facilitate the application of self-consistent expansions to other dynamic systems, and illustrate its application using the example of Barenblatt's equation.  
%revision3.1
The scope of our results on the application of renormalization group and self-consistent expansions is limited to partial differential equations whose long-time asymptotics is controlled by incomplete similarity.  
However, our work suggests that these methods could be applied to a broader suite of singular perturbation problems such as boundary layer theory, multiple scales analysis and matched asymptotic expansions, for which excellent approximations using renormalization group methods alone are already available.

\end{abstract}

\maketitle
\end{CJK*}
%\tableofcontents
\section{\label{sec:level1}Introduction}
% Physics problem as perturbation theory
For most physical systems, exact solutions are  unattainable.
Perturbation theory is a practical approximation approach that treats the full system as a small deviation from an exactly solvable base system. 
By expanding in powers of the small perturbation and solving corrections to the base system order by order, the solution is constructed as a series in the perturbation strength.

% Singular perturbation and resummation methods
Unfortunately, naive perturbation theory encounters limitations in numerous scenarios, in particular those where individual terms in the expansion are divergent as an asymptotic limit is taken (for example, long time) \cite{cole1996self,nayfeh2008perturbation,van1975perturbation,bender1999advanced,murdock1999perturbations,kuehn2015multiple}.  
In addition, even if the individual terms are finite, the perturbation series may itself be divergent \cite{bender1969anharmonic,bender1971large,dingle1973asymptotic,boyd1999devil,boyd2005hyperasymptotics,berry1990hyperasymptotics}.  
To address these challenges, a variety of methods have been developed by applied mathematicians, engineers and physicists that effectively resum the perturbative solution, thus removing divergences and enabling the accurate determination of asymptotic forms.  
A necessarily brief list of examples where these methods have found success include renormalization and the short distance behavior of quantum field theory (specifically quantum electrodynamics), where perturbative renormalization group (RG) methods were originally invented and deployed \cite{stueckelberg1951normalization,stueckelberg1953normalization,gell1954quantum,zinn2021quantum}, and critical behavior at phase transitions, where statistical field theory predicts anomalous dimensions and departures from mean field theory exponents \cite{larkin1969phase,wilson1971renormalization,wilson1972critical,goldenfeld2018lectures}.  
These field theory examples all exhibit scale invariance, but more recently it was recognized that partial differential equations can also exhibit anomalous dimensions arising in scale-invariant similarity solutions that cannot be constructed by elementary methods related to dimensional analysis and simple analyticity assumptions \cite{barenblatt1972self,BarenblattSivashinskii1969,barenblatt1996scaling}; renormalization group methods have been successfully applied here too \cite{barenblatt1996scaling,goldenfeld1989intermediate,goldenfeld1990anomalous,chen1992renormalization,bricmont1992renormalization,bricmont1994renormalization,goldenfeld2018lectures}.  
Finally, we mention more complex deterministic differential equation problems with multiple scales and boundary layers arising in fluid mechanics and other areas of continuum mechanics, where matched asymptotic expansions, singular perturbation theory and even renormalization group methods have been used~\cite{bender1999advanced, goldenfeld2018lectures,chen1996renormalization,veysey2007simple,kuehn2015multiple,boyd1999devil,boyd2005hyperasymptotics,ei2000renormalization,kirkinis2012renormalization,galvez2021numerical,antonov2002field} and problems where perturbation theory is itself unable to account for intrinsically non-perturbative singular phenomena, where methods based on exponential asymptotics beyond all orders \cite{berry1990hyperasymptotics,Berry1991,dingle1973asymptotic,boyd2005hyperasymptotics} have been developed.  %revision1
Amongst the approximation methods capable of handling difficult perturbation problems, we mention large-order perturbation theory~\cite{mera2015nonperturbative,zinn2021quantum}, exact renormalization group~\cite{morris1994exact,bagnuls2001exact,berges2002non,mera2015nonperturbative, rosten2012fundamentals}, and resurgence theory~\cite{ecalle1981fonctions,aniceto2019primer}.  Despite the specific successes of these and other methods, there is still a need for simple but accurate methods that work even for strong coupling problems, and are capable of detecting subtle singular features that are hard to capture by existing techniques.  

% Self-consistent theory
%In this paper, we develop such an approach, bringing together perturbative renormalization group methods 
%Revision3.2
In this paper, we develop such an approach, bringing together the renormalization group framework  
\cite{goldenfeld1989intermediate,goldenfeld1990anomalous,chen1992renormalization,bricmont1992renormalization,bricmont1994renormalization,goldenfeld2018lectures} with a lesser-known approximation method known as the self-consistent expansion (SCE) that has in one form or another been rediscovered several times in different contexts, perhaps beginning with the work of Edwards, Herring, Phythian and Kraichnan in turbulence theory \cite{edwards1964statistical,herring1965self,phythian1969self,kraichnan1970convergents,sreenivasan2005sam}, several works in the context of interface dynamics and other fields,\cite{schwartz1992nonlinear,bouchaud1993self,prakash1994long,schwartz1998peierls,cohen2016self,schwartz2008ideas} and, as we will mention below, arguably in Bogoliubov's work on interacting Bose gases \cite{bogoliubov1947theory}.  
We will introduce this approach to approximation theory below, but for now, we simply mention that self-consistent expansions have been shown to be unexpectedly effective in tackling strong coupling problems.  
Thus, it is a natural question to ask whether or not the self-consistent expansion method can be applied to problems with anomalous dimensions, thus improving the accuracy of methods such as the $\epsilon$-expansion in renormalization group theory \cite{wilson1972critical}.  
%The new ingredient added here is to combine this approach with perturbative renormalization group  methods, in order to further improve the RG approximant for the singular long-time behaviour of non-linear partial differential equations.  
%Revision3.3 
The new ingredient added here is to implement the SCE method starting with a variational form inspired by RG methods, in order to further improve the RG approximant for the singular long-time behaviour of non-linear partial differential equations.  

The central strategy of the self-consistent expansion is to impose ad hoc constraints on the system, requiring it to adhere to certain symmetries or closure approximations through self-consistent rules. 
The intent is to enable a low order approximation to incorporate information that is either higher order in perturbation theory or even beyond the reach of perturbation theory.  
For example, in some implementations of the self-consistent approximation, a perturbation theory is developed, and then an ad hoc condition is imposed that the coefficient of the second order term in the expansion is forced to be zero, which in turn causes the zeroth order solution to be determined in a way that can even be non-analytic in a coupling parameter.  
Such self-consistent expansions can also be thought of as variational methods, because they can be justified as implementing some sort of principle of minimum sensitivity \cite{stevenson1981optimized,yukalov2019interplay,stevenson2022renormalized} or even a form of renormalization group, where one endeavours to improve renormalized perturbation theory by minimizing the dependence on the renormalization scale \cite{goldenfeld2018lectures}.  
Another recent approach is \lq\lq iterative perturbation theory" \cite{smerlak2022perturbation}, where the partitioning of the Hamiltonian strategy is carried out and examined at high order.

Perhaps the earliest example of this sort of strategy is the  operator transformation that was used by Bogoliubov in his theory of the weakly interacting Bose gas \cite{bogoliubov1947theory,andersen2004theory,goldenfeld2023there}.  Here, the expansion of the Hamiltonian in terms of the strength of the scattering between Bosons leads to anomalous terms at second order that cannot be easily interpreted; but setting their coefficient to zero, leads to a unique determination of the zeroth order term in the expansion.  The resulting excitation spectrum is a collective outcome of the real interacting Bosons in terms of non-interacting quasi-particles (in reality higher order terms lead to weak interactions and finite lifetimes to these quasi-particles \cite{andersen2004theory}).  This description predicts that the ideal Boson condensate at zero temperature is depleted due to the residual interactions in a way that is non-analytic in terms of the interaction strength.  Such a result could not be obtained by a pure perturbation calculation, which would by necessity lead to an analytic formula.  

These methods, due to their variational nature, are generally more straightforward in terms of calculation. However, they can be less transparent in the construction of the problem. 
In this paper, we focus on the self-consistent expansion (SCE) method, which integrates perturbation theory with the idea of self-consistency. 
This combination leverages the practical implementation and interpretability of perturbation theory and the variational flexibility of self-consistency constraints.
Within the framework of perturbation theory, the fundamental concept of the SCE method involves introducing variational parameters to re-partition the unperturbed and perturbed components of a system. 
These parameters are then optimized to ensure that the adjusted perturbation theory remains ``self-consistent'' at the targeted order of approximation. 
Typically, the self-consistent criterion, or optimization condition, involves selecting a key physical quantity of interest and ensuring that its next-order correction vanishes under the new perturbation theory.

% History of SCE and current status
A remarkably simple form of the SCE method was developed by Edwards and Singh in 1979 as a ``rapid and accurate" approach to tackle the self-avoiding random walk problem in polymers~\cite{edwards1979size}; we will briefly review this work in the next section. More recently, the SCE method and related methods have been extended to singular physics problems such as the Stark effect \cite{mera2015nonperturbative}, quantum anharmonic oscillator \cite{mera2015nonperturbative,remez2018divergent} and the Kardar-Parisi-Zhang model of interface growth \cite{bouchaud1993self,katzav1999self,remez2018divergent,schwartz1992nonlinear,schwartz1998peierls}, as well as to mathematical problems such as asymptotic expansions of special functions~\cite{remez2018divergent, schwartz2008ideas} and turbulence modeling \cite{eyink1998predictive}. 
%
%
% We apply SCE to Barenblatt's equation; provide a field-theoretic framework
%In previous studies, the SCE has been applied to equilibrium problems in quantum and statistical physics, non-equilibrium problems in statistical physics, and to a lesser extent, to asymptotics problems.

The purpose of this paper is to extend the scope of the SCE to deterministic, spatially-extended dynamical systems with anomalous dimensions. 
As a first step, we show how SCE methods can improve the RG result of time-dependent singular perturbation problems, using, as an example, flow in porous media governed by Barenblatt's equation~\cite{barenblatt1996scaling}. Methodology-wise, the SCE method has been primarily applied to problems expressible in an integral form, where the action is simply a linear sum of the unperturbed part and the perturbation term, facilitating a straightforward re-partitioning step. This framework makes the application of SCE very similar to a standard field theory calculation of perturbation theory. Schwartz and Katzav have shown how to apply SCE to stochastic nonlinear field theory using the Fokker-Planck equation~\cite{schwartz2008ideas,remez2018divergent}. Here, we provide a field-theoretic framework for the deterministic dynamical model of Barenblatt's using the Martin-Siggia-Rose (MSR) formalism~\cite{martin1973statistical} and apply SCE to this alternative form as well. We remark that the MSR technique has previously been used to formulate an action for Barenblatt's equation, enabling the use of exact RG to obtain the known asymptotic result perturbatively~\cite{yoshida2005exact}.

% Structure of the paper
Our paper is structured as follows: Section~\ref{sec:Polymer} is a brief review of the  SCE procedure to solve the problem of a single polymer chain in solution with excluded volume. Section~\ref{sec:PME} introduces Barenblatt's equation describing groundwater spreading in porous medium as the physical problem to which we will apply the SCE method. We review how usual perturbation theory leads to a divergent expansion and how perturbative RG identifies the correct self-similar form. Next in Sec.~\ref{sec:SCE}, we show that SCE improves results from perturbative RG, particularly in the large perturbation regime. While all the calculations above are within the conventional framework of solving PDEs with Green's functions, we shift our focus in Sec.~\ref{sec:FT} to transforming the dynamical system problem into a field-theoretic form using the MSR formalism. We show that equivalent RG and SCE results can be achieved in a field-theoretic context, broadening the applicability of these methods. Finally in Sec.~\ref{sec:conclusion}, we discuss potential avenues for future work, including extensions of SCE to other problems and computational frameworks.

\section{\label{sec:Polymer}Review of the SCE calculation for Polymer problem}
In this section, we briefly review the instructive SCE calculation for the polymer self-avoiding walk~\cite{edwards1979size}. A polymer chain is modeled as a continuous path that avoids itself due to the excluded volume effect, and its statistical properties can be studied using the Edwards measure~\cite{edwards1965statistical}: 
\begin{align}\label{wiener}
    \rho[\textbf{r}(s)] &= e^{-\frac{3}{2l}\int_0^L \dot{\textbf{r}}^2(s) ds - \omega\int_0^L\!\int_0^L\delta\left[\textbf{r}(s)-\textbf{r}(s') \right] ds ds'}  \nn\\
    &\equiv e^{-S_0 (l) - S_\omega},
\end{align}
where $L$ is the total length of a polymer chain, composed of $N$ segments of step length $l$, $s$ is the arc length, $\textbf{r}$ is the spatial coordinate and $\omega$ is the strength of the interaction potential. This measure can be interpreted as a perturbation theory in a path integral form, where the zeroth-order system is described by the Wiener measure of an ordinary random walk, and the self-exclusion interaction serves as a perturbation.  Strictly speaking, the delta-function self-interaction in the path integral is a pseudo-potential in the same spirit as the pseudo-potential used in Bogoliubov's theory of the weakly-interacting Bose gas \cite{bogoliubov1947theory,goldenfeld2023there}.  The Edwards-Singh approach has been extended for general potentials \cite{raphael1992one}, although it is not clear to us whether the calculation as presented is applicable beyond the pseudo-potential approximation.  This may be relevant for the puzzling technical results reported when Coulomb potentials are used (see Discussion Sec. 1 of Ref.~(\cite{raphael1992one})).

To quantify the average size of polymer chains, we want to calculate the moment of the end-to-end distance, denoted by $R$, in the framework of the Edwards model 
\begin{align}\label{R_dist}
    \left\langle R^2 \right\rangle = \frac{\int D\textbf{r}\left[\textbf{r}(L) - \textbf{r}(0)\right]^2\rho[\textbf{r}]}{\int D\textbf{r}\,\rho[\textbf{r}]}.
\end{align}
Here, the self-excluding interaction as a strong coupling makes a regular perturbation expansion divergent, which can be resummed using renormalization group (RG) method~\cite{de1972exponents,oono1985statistical}. Edwards and Singh, however, took a different approach: they reorganized the action by introducing a variational length scale $l_1$ such that $\left\langle R^2 \right\rangle = Ll_1$:
\begin{align}
    \rho[\textbf{r}] = \exp\left\{-S_0 (l_1) - \left[S_0(l) -S_0(l_1) + S_\omega\right]\right\},
\end{align}
where the rescaled unperturbed term is $S_0 (l_1) $ and the new perturbation is $S' \equiv \left[S_0(l) -S_0(l_1) + S_\omega\right]$.  The physical meaning of $l_1$ is that it is an effective or renormalized step length for a fictitious ideal Brownian chain, which takes into account the interactions.  Thus, in the same way that a weakly-interacting Bose gas can be represented as a gas of effective non-interacting Bosons, i.e. quasi-particles whose dispersion relation reflects the actual interactions between the Bose gas atoms, and thus is not that of ideal Bosons, so the interacting polymer chain is represented as an effective non-interacting Brownian chain but with a renormalized step length that depends on the interactions. 

Next, they conduct a standard perturbation expansion based on the new perturbation theory:
\begin{align}\label{newPT}
    \left\langle R^2 \right\rangle = \frac{\int D\textbf{r}\left[\textbf{r}(L) - \textbf{r}(0)\right]^2\exp\left[-S_0 (l_1)\right]\left[1 - S' + \text{(h.o.t.)}\right]} {\int D\textbf{r}\,\exp\left[-S_0 (l_1)\right]\left[1 - S' + \text{(h.o.t.)}\right]}.
\end{align}
The self-consistent criterion here is to set the first-order correction to $\left\langle R^2 \right\rangle $ to zero, and therefore determine the $l_1$:
\begin{align}
    Ll_1^2 \left(\frac{1}{l} - \frac{1}{l_1}\right) = C_0\omega\frac{L^{3/2}}{l_1^{1/2}},
\end{align}
where the constant $C_0 = 2\sqrt{\frac{6}{\pi^3}}$. There are two asymptotic regimes for $l_1$.  First there is a regime where perturbation theory (PT) is adequate, and second, there is a strong interaction regime where the effective step length is strongly renormalized by the interactions:
\begin{equation}
    l_1 = \left\{
        \begin{array}{ll}
            l + C_0\omega L^{1/2} l^{-1/2}, & l_1 \approx l \;\text{(PT regime)} \\
            (C_0\omega l)^{2/5} L^{1/5},  & l_1 \gg l
        \end{array}
    \right. .
\end{equation}
The latter regime $l_1 \gg l$ corresponds to the asymptotic limit
$L\to\infty$. Therefore,
\begin{equation}
\left\langle R^2 \right\rangle = Ll_1 = (C_0\omega l)^{2/5} L^{6/5} \propto L^{6/5}, \quad\text{as}\; L\to\infty.
\end{equation}
This short calculation yielded remarkably accurate results for the anomalous dimension, recovering the Flory exponent $\alpha = 6/5$.  This is known to be a very good approximation to other results obtained by RG or numerical simulation: $\left\langle R^2 \right\rangle \propto L^{\alpha}$, where  $\alpha =1.195$~\cite{de1972exponents} and the numerical result $\alpha = 2\mu, \mu = 0.58759700(40)$\cite{clisby2016high}. 

We introduce this example because it has many analogies to the approach we will take in this paper to the problem of anomalous dimensions in partial differential equations.  The self-avoiding walk for an isolated polymer chain in solution was originally formulated as a path integral, and then mapped into a partial differential equation framework by Edwards~\cite{edwards1965statistical,de1969polymer,freed1972functional}, using a self-consistent field closure.  In this approach, the arc length coordinate along the polymer chain is equivalent to time, whereas the position of the polymer is the space coordinate, and the equation for the propagator of the polymer chain obeys a Green function equation, something analogous to the Schwinger-Dyson equation in quantum field theory. This narrative reveals a clear parallel to the core subject of this paper, Barenblatt's equation \cite{BarenblattSivashinskii1969,barenblatt1972self,barenblatt1996scaling,goldenfeld1989intermediate,goldenfeld1990anomalous,goldenfeld2018lectures}, but working backwards.  Specifically, we consider a partial differential equation, whose solution was originally formulated directly using Green's functions and an integral equation \cite{goldenfeld1989intermediate,goldenfeld1990anomalous,goldenfeld2018lectures} and solved using perturbative RG, but which in the present paper, we formulate in terms of a path integral framework. In both these formulations, we apply the self-consistent expansion in conjunction with RG, and show the equivalence of the results.

\section{\label{sec:PME}Barenblatt's equation: a singular perturbation problem}

Barenblatt proposed a nonlinear diffusion equation that models the flow of ground water in an elasto-plastic porous medium (i.e. a sponge!)~\cite{barenblatt1996scaling}:
\begin{align}  \label{eq:barenblatt}
\partial_t u &= \frac{\kappa}{2} \left[1 + \epsilon\,\Theta(-\partial_t u)\right] \partial_x^2 u \nn\\
u(x, 0) &\equiv v_0 (x) = \frac{Q_0}{\sqrt{2\pi l^2}} \exp\left(-\frac{x^2}{2 l^2}\right).
\end{align}

Formal solution through the Green's function method yields a Volterra integro-differential equation that can be solved iteratively: 
\begin{multline}
u(x, t) = \int_\mathbb{R} dy \,G_0 (x, y; t, 0) \,v_0(y) \\
+ \frac{\epsilon\kappa}{2} \int_0^t ds\int_{-X_\epsilon}^{X_\epsilon} dy\,G_0 (x, y; t, s) \,\partial_y^2 u(y,s), \label{eq:formal_solu}
\end{multline}
where the integral limit $X_\epsilon (t)$ is given by
\begin{equation}
\partial_t u(x, t) = 0 \big|_{|x| = X_\epsilon (t)},
\end{equation}
and $G_0$ is the usual Green's function for diffusion equation
\begin{equation}\label{green}
G_0(x, y; t, s) = \frac{1}{\sqrt{2\pi\kappa (t-s)}} \exp\left[-\frac{(x-y)^2}{2 \kappa (t-s)}\right].
\end{equation}
The presence of $u(x,t)$ on both sides of Eq.~(\ref{eq:formal_solu}) make this a challenging mathematical problem to solve analytically, although in this particular case it is possible to relate the solutions to special functions \cite{BarenblattSivashinskii1969} by making the ansatz that there are anomalous scaling exponents due to incomplete similarity \cite{barenblatt1972self,barenblatt1996scaling}. A brute force attack on the problem is possible using perturbation theory in the small parameter $\epsilon$ limit, where the zeroth order system is the linear diffusion, and the discontinuity in the diffusion coefficient gives the perturbation term from which a RG calculation yields a perturbation expansion in $\epsilon$ for the anomalous dimensions \cite{goldenfeld1989intermediate,goldenfeld1990anomalous,goldenfeld2018lectures}.  This approach is of course purely formal, but has been made rigorous \cite{kamin1991barenblatt,aronson1994calculation} including results on the existence and behavior of the anomalous dimensions as a function of $\epsilon$ \cite{kamin1991barenblatt}.

\subsection{\label{sec:PT} Divergence of the usual perturbation theory}
In the usual perturbation theory, we expand the solution around a small parameter $\epsilon$,
\begin{align}
u(x, t) &= u_0(x,t) + \epsilon\, u_1(x,t) + \cdots \nn\\
X_\epsilon (t) &= X_0 (t) + \epsilon X_1 (t) + \cdots ,
\end{align}
and plug the expansions back into Eq.~(\ref{eq:formal_solu}) to obtain:
\begin{multline}
u(x, t) = \int_\mathbb{R} dy \,G_0 (x, y; t, 0) \,v_0(y) \\
+ \frac{\epsilon\kappa}{2} \int_0^t ds\int_{-X_0}^{X_0} dy\,G_0 (x, y; t, s) \,\partial_y^2 u_0(y,s) + \mathcal{O}(\epsilon^2). \label{eq:naive_expansion}
\end{multline}
Solving to order  $\epsilon$, we obtain the zeroth order solution and the first order correction
\begin{align}\label{psolu}
u_0(x, t) &= \int_\mathbb{R} dy \,G_0 (x, y; t, 0) \,v_0(y) = \frac{Q_0 \, e^{- x^2 / \left[2 (\kappa t + l^2)\right]}}{\sqrt{2\pi (\kappa t + l^2)}} \nonumber\\
u_1(x, t) &= -\frac{1}{\sqrt{2\pi e}}\ln\left(\frac{\kappa t + l^2}{l^2}\right) u_0  + \left(\text{r.t.}\right),
\end{align}
where ``r.t." represents ``regular terms". In the limit $\frac{\kappa t}{l^2}\to\infty $, 
\begin{multline}\label{naivePT}
    u(x, t) \sim \frac{Q_0\, e^{-x^2 / (2\kappa t)}}{\sqrt{2\pi\kappa t}} \left[1 -\frac{\epsilon}{\sqrt{2\pi e}}\ln\left(\frac{\kappa t}{l^2}\right)\right] \\
    + \epsilon \cdot \left(\text{r.t.}\right) + \mathcal{O}(\epsilon^2),
\end{multline}
where the first order term contains a logarithmic divergence, breaking the self-similarity at long-times, showing that the solution retains a long-time memory of the initial condition. However, this secular term is an artifact of the naive perturbation theory because a bounded solution to the Cauchy problem exists~\cite{kamenomostskaya1957problem,kamin1991barenblatt,barenblatt1996scaling}. In short, with its unusual feature of a discontinuous diffusion coefficient (as a function of whether or not $u$ is increasing or decreasing) Barenblatt's equation may be regarded as a singular perturbation problem and we need new methods to find the correct asymptotic form.

\subsection{The origin of divergence}
To find the origin of the logarithmic divergence in perturbation theory, we first perform dimensional analysis, a powerful tool to analyze the behaviors of a physical system with scaling laws and self-similarity. In this problem, we have physical quantities with the following units:
\begin{gather}
[Q_0] = M, \, [u] = M/L, \, [x] = [l] = L, \, [t] = T, \nonumber \\
[\kappa] = L^2 T^{-1}, [\epsilon] = 1.
\end{gather}
We construct the corresponding dimensionless quantities and rewrite the solution in terms of dimensionless quantities:
\begin{gather}
\Pi = \frac{\sqrt{\kappa t}}{Q_0} u, \, \Pi_1 = \frac{x}{\sqrt{\kappa t}},  \, \Pi_2 = \frac{l}{\sqrt{\kappa t}},  \, \Pi_3 = \epsilon, \\
\Pi = f \, (\Pi_1, \Pi_2, \Pi_3) \,\Rightarrow \, u = \frac{Q_0}{\sqrt{\kappa t}} f \, \left(\frac{x}{\sqrt{\kappa t}}, \frac{l}{\sqrt{\kappa t}}, \epsilon\right),
\end{gather}
where the perturbation expansion of the dimensionless function $f$ around $\epsilon$ is secular at the limit $\Pi_2 = \frac{\kappa t}{l^2}\to\infty$. This singularity is precisely the origin of the logarithmic divergence in the naive perturbation theory. While this breaks the self-similarity deduced from dimensional analysis (namely intermediate asymptotics of the first kind), Barenblatt's equation retains a nontrivial form of self-similarity, categorized as intermediate asymptotics of the second kind arising from incomplete similarity in the limit $\Pi_2 = \frac{\kappa t}{l^2}\to\infty$ \cite{barenblatt1972self,barenblatt1996scaling, goldenfeld2018lectures}.

\subsection{Self-similar solution at the long-time asymptotic regime}\label{subsec:self_similar}
The asymptotic regime of concern is $\frac{\kappa t}{l^2}\to\infty $, which can be achieved equivalently by taking the limit of long time $t$ with finite $l$, or by fixing the time $t$ and taking $l \to 0$. We choose the latter procedure in the rest of this calculation, but all the methods are of course applicable to the former one.

To renormalize the perturbation theory, we introduce an arbitrary finite length scale $\mu$ where $[\mu] = L$ and therefore the quantity $\frac{\mu}{\sqrt{\kappa t}}$ remains finite. We define the renormalized function $F$ and the renormalization factor $Z$, and choose $Z$ to eliminate the leading order of divergence
\begin{align}
F\left(\frac{x}{\sqrt{\kappa t}}, \frac{\mu}{\sqrt{\kappa t}}, \epsilon\right) = Z\left(\frac{l}{\mu}\right) \cdot f\left(\frac{x}{\sqrt{\kappa t}}, \frac{l}{\sqrt{\kappa t}}, \epsilon\right).
\end{align}
Because the bare function $f$ is independent of the arbitrary length scale $\mu$, we obtain the RG equation
\begin{align}\label{PDE_CS}
\mu \frac{d f}{d \mu} = 0 \quad
&\Rightarrow \,  \mu \frac{d(Z^{-1} F)}{d \mu} = 0 \quad \\
&\Rightarrow \,  \left[-\frac{d \ln Z}{d \ln \mu} + \sigma \frac{\partial}{\partial \sigma}\right] F = 0,
\end{align}
where $\sigma \equiv \frac{\mu}{\sqrt{\kappa t}}$, and the last equation is reminiscent of the Callan-Symanzik equation in quantum field theory ~\cite{callan1970broken,symanzik1970small,zinn2021quantum}. Later in this paper we will show that this is no accident: it is precisely the Callan-Symanzik equation when expressed in the field-theoretic framework \cite{callan1970broken,symanzik1970small,zinn2021quantum,goldenfeld2018lectures}.

In the limit $l \to 0$, we impose the renormalizability assumption
\begin{align}\label{eq:renorm_assp}
\lim_{l \to 0} \frac{d \ln Z (l/\mu) }{d \ln \mu} = \text{dimensionless constant} \equiv 2 \alpha ,
\end{align}
and solve Eq.~(\ref{PDE_CS}) to find the self-similar form
\begin{align} \label{PDE_SS}
u = \frac{Q_0}{\sqrt{\kappa t}} \left(\frac{l}{\sqrt{\kappa t}}\right)^{2\alpha}\varphi \left(\frac{x}{\sqrt{\kappa t}}, \epsilon\right) \propto t^{- \frac{1}{2} - \alpha} \, \varphi \left(\frac{x}{\sqrt{\kappa t}}, \epsilon\right).
\end{align}

In summary, requiring there to be a self-similar solution at long time for the renormalized perturbation theory means that the solution must involve an anomalous dimension which enters in the form of Eq.~(\ref{PDE_SS}).  In Sec.~\ref{sec:SCE}, we will combine this constraint with the self-consistent expansion to generate an approximation that cannot be obtained by either the original RG method or the self-consistent expansion method alone.

\subsection{Perturbative RG}\label{subsec:PDE_RG}
The RG equation gives a self-similar solution in the asymptotic regime $l \to 0$, but the value of the anomalous dimension $\alpha$ is unknown. A perturbative RG calculation \cite{goldenfeld1989intermediate,goldenfeld1990anomalous} (reviewed in pedagogical detail in Chapter 10 of Goldenfeld's textbook~\cite{goldenfeld2018lectures}) approximates $\alpha$ by renormalizing $Q$ in the perturbation expansion. Here, we omit the calculation details but give the final results directly. The final result of the first order perturbative RG is, as $\frac{\kappa t}{l^2} \to \infty$,

\begin{equation}\label{perRG_solu}
u(x, t) \sim \frac{Q_0}{\sqrt{2\pi\kappa t}} \left( \frac{\kappa t}{l^2}\right)^{-\alpha} \exp\left(-\frac{x^2}{2\kappa t}\right) + \mathcal{O}(\epsilon^2),
\end{equation}
where the anomalous dimension is $\alpha = \frac{\epsilon}{\sqrt{2\pi e}} + \mathcal{O}(\epsilon^2)$. Extension of these results to higher order was achieved by Cole \el \cite{cole1996self} who used a Lie group method to obtain 
\begin{equation}
    \alpha(\epsilon) = \frac{\epsilon}{\sqrt{2\pi e}} - 0.063546\,\epsilon^2 + \mathcal{O}(\epsilon^3),
\end{equation}
a result also obtained by Yoshida \el using the so-called exact renormalization group method.

While these methods provide highly accurate perturbative approximations of the asymptotics, they involve complex and lengthy mathematical derivations.

\section{\label{sec:SCE}Application of self-consistent expansion to Barenblatt's equation}
In this section, we will apply the SCE method to the Barenblatt's equation to find the long-time asymptotics and compare the results to those from other methods.
\subsection{SCE in PDE framework}\label{subsec:SCE_PDE}
We start with Eq.~(\ref{eq:formal_solu}), an integro-differential equation derived in the PDE framework using the Green's function method. The first step is to rescale the zeroth order system to obtain a new perturbation theory. In ordinary perturbation theory, the zeroth order equation is 
\begin{align}
u_0 &= \frac{Q_0}{\sqrt{2\pi (\kappa t + l^2)}} \exp\left[-\frac{x^2}{2(\kappa t+l^2)}\right] \nonumber\\
&\sim \frac{Q_0}{\sqrt{2\pi\kappa t}} \exp\left(-\frac{x^2}{2\kappa t}\right),  \quad \text{as} \; \frac{\kappa t}{l^2} \to \infty,
\end{align}
which is of the self-similar form Eq.~(\ref{PDE_SS}) when $\epsilon = 0$, resulting in $\alpha = 0$. 

To proceed, we use the solution Eq.~(\ref{PDE_SS}) of the Callan-Symanzik equation~(\ref{PDE_CS}) with $\alpha\neq 0$ to construct the perturbation theory for long times. We can further simplify Eq.~(\ref{PDE_SS}) by expanding the function $\varphi$, 
which is regular in the limit of $\frac{\kappa t}{l^2} \to \infty$, 
with respect to $\epsilon$ and keep only the leading order in  $\epsilon$:
\begin{align}
u &\sim \frac{Q_0}{\sqrt{\kappa t}} \left(\frac{l}{\sqrt{\kappa t}}\right)^{2\alpha} \left[ \varphi_0 \left(\frac{x}{\sqrt{\kappa t}}\right) + \mathcal{O}(\epsilon) \right] \nonumber\\
&\sim \frac{Q_0}{\sqrt{2\pi\kappa t}} \left(\frac{l}{\sqrt{\kappa t}}\right)^{2\alpha}  \exp\left(-\frac{x^2}{2\kappa t}\right).
\end{align}
When $\alpha = 0$ the long-time asymptotics of the self-similar solution aligns with the long-time asymptotics of the original zeroth order system, so we choose it as the zeroth order system in the new perturbation theory. To match with the initial condition at $t = 0$ and the boundary condition at $\epsilon = 0$, 
we write the self-consistent solution in the following form
\begin{equation}
u_\alpha = \frac{Q_0}{\sqrt{2\pi (\kappa t + l^2)}} \left(\frac{l}{\sqrt{\kappa t + l^2}}\right)^{2\alpha}  \exp\left[-\frac{x^2}{2(\kappa t+l^2)}\right],
\end{equation}
where $\alpha$ represents a variational parameter to be determined by a self-consistent criterion, and may depend on other physical quantities in the problem. Here $\alpha$ can be a function of $\epsilon$ and must satisfy $\alpha (\epsilon = 0) = 0$ as a boundary condition.

Now we construct a new perturbation theory where we re-partition Eq.~(\ref{eq:formal_solu}) with the new unperturbed term $u_\alpha$ and define the corresponding new perturbation term (PT) $u_\alpha^I$:
\begin{align}
u(x, t) &= u_0 + u_\epsilon^I \quad \text{(old PT)} \nonumber\\
        &= u_\alpha + \left( -u_\alpha + u_0 + u_\epsilon^I \right) \\
        &\equiv u_\alpha + u_\alpha^I  \quad \text{(new PT)} ,\nonumber 
\end{align}
where the new perturbation term is 
\begin{align}
u_\alpha^I &\equiv - u_\alpha + u_0  + \frac{\epsilon\kappa}{2} \int_0^t ds\int_{-X_\epsilon}^{X_\epsilon} dy\,G_0 (x, y; t, s) \,\partial_y^2 u(y,s) \nn\\
&\equiv u_\alpha^{(1)} + u_\alpha^{(2)} + \cdots.
\end{align}
The leading order correction becomes 
\begin{multline}\label{SCE_u1}
u_\alpha^{(1)} = -u_\alpha + u_0 \\
+ \frac{\epsilon\kappa}{2} \int_0^t ds\int_{-X_\alpha (s)}^{X_\alpha (s)} dy\,G_0 (x, y; t, s) \,\partial_y^2 u_\alpha(y,s), 
\end{multline}
where $X_\alpha (t) = \sqrt{(1+ 2\alpha)(\kappa t + l^2)}$ should be solved from
\begin{align}
\partial_t u_\alpha(x, t) = 0 \big|_{|x| = X_\alpha (t)} 
\end{align}

Next, we impose the self-consistent criterion at first order in the SCE to solve for $\alpha$. Here, we choose the total mass $m(t) = \int_\mathbb{R} dx\, u(x, t) $ as the quantity of physical significance to establish this criterion, which means its first-order correction should vanish under our new perturbation theory. This is because at the current order, the essential physics we are trying to capture is the adiabatic loss of mass over time. When $\epsilon = 0$, mass is conserved, i.e. $m_0(t) = Q_0$; when $\epsilon > 0$, we find at the asymptotic limit $\frac{\kappa t}{l^2}\to \infty$, the dominant contribution of the perturbation to $u(x,t)$ is the renormalization factor $\left(l/\sqrt{\kappa t+l^2}\right)^{2\alpha}$, which is independent of position $x$. Therefore using the total mass $m$ should be a natural choice for estimating $\alpha$. Under the new perturbation theory, we find the perturbative expansion of the total mass
\begin{align}
m(t) = \int_\mathbb{R} dx\, u(x, t) = \int_\mathbb{R} dx\, \left[ u_\alpha + u_\alpha^{(1)} + u_\alpha^{(2)} + \cdots \right],
\end{align}
and the self-consistent criterion sets 
\begin{widetext}
\begin{align}
0 \stackrel{SCE}{\equiv} \int_\mathbb{R} dx\, u_\alpha^{(1)}  &= \int_\mathbb{R} dx \left[-u_\alpha + u_0 + \frac{\epsilon\kappa}{2} \int_0^t ds\int_{-X_\alpha (s)}^{X_\alpha (s)} dy\,G_0 (x, y; t, s) \,\partial_y^2 u_\alpha(y,s)\right] \\
   &= Q_0 \left[1 - \left(\frac{l}{\sqrt{\kappa t+ l^2}}\right)^{2\alpha}\right] + \frac{\epsilon\kappa}{2}  \int_0^t ds\int_{-\sqrt{(1+ 2\alpha)(\kappa s + l^2)}}^{\sqrt{(1+ 2\alpha)(\kappa s + l^2)}} dy\, \frac{Q_0 l^{2\alpha}}{\sqrt{2\pi} (\kappa s + l^2)^{\frac{3}{2} + \alpha} }\left(\frac{y^2}{\kappa s + l^2}-1\right) e^{-\frac{y^2}{2(\kappa s + l^2)}} \nonumber\\
   &= Q_0 \left[1 - \left(\frac{l}{\sqrt{\kappa t+ l^2}}\right)^{2\alpha}\right] + \frac{\epsilon\,Q_0 } {2\sqrt{2\pi}} \left(-\frac{1}{\alpha}\right)\left[\left(\frac{l}{\sqrt{\kappa t+ l^2}}\right)^{2\alpha} -1 \right]\left(-2\sqrt{1+2\alpha} \, e^{-\frac{1}{2} -\alpha}\right),
\end{align}
\end{widetext}
where the condition $\alpha\neq 0$ is used in the time integral to yield a non-trivial result; when $\alpha = 0$, it leads to a logarithmic function in time, rather than the power law derived above. Recall that $\alpha = 0$ corresponds to the unperturbed case ($\epsilon = 0$), so the calculation above reveals the exact origin of the logarithmic divergence in perturbation theory, where the unperturbed solution $u_0$ is integrated over in the first order calculation.
We collect terms and find the self-consistent parameter $\alpha$ is determined by the following equation
\begin{gather}
\left[1 - \left(\frac{l}{\sqrt{\kappa t+ l^2}}\right)^{2\alpha}\right] \left(1 - \frac{\epsilon } {\sqrt{2\pi e}} \frac{\sqrt{1+2\alpha} }{\alpha} e^{-\alpha}\right) = 0 \nonumber \\
\Rightarrow \, \alpha = \frac{\epsilon } {\sqrt{2\pi e}} \sqrt{1+2\alpha} \, e^{-\alpha},
\label{eq:final_result}
\end{gather}
which we solve numerically, with the results illustrated by the red dashed line in Fig.~\ref{fig:comparison}.  We emphasize that in obtaining Eq.~(\ref{eq:final_result}), we had to use both the Callan-Symanzik equation~(\ref{PDE_CS}) from renormalization group and the self-consistent expansion. We have not been able to obtain this result from either the RG or SCE methods alone.
Note the SCE method at the first order yields the same results as an iterative method~\citep{chen1998anomalous} because we apply SCE directly to the integro-differential equation~(\ref{eq:formal_solu}), which is also the self-referential formal solution of the targeted quantity $u(x, t)$. 
In Sec.~\ref{sec:FT}, we will introduce a new framework as an effort to separate the local interaction (e.g., a Hamiltonian) and the targeted physical quantity.

\subsection{Comparison of methods}
We compare our value of $\alpha$ obtained from the SCE method to the results of RG and the exact  solution. Starting with the self-similar form, the exact values of the anomalous dimension $\alpha (\epsilon)$ and the factor $\xi_{\epsilon}$ can be found by solving the following transcendental equations~\cite{barenblatt1996scaling,goldenfeld2018lectures}:
\begin{gather}
 D_{2\alpha + 2} \left(\xi_\epsilon\right) = 0, \label{exact_anom1} \\
 F\left(-\alpha.-1, 1/2, \xi_{\epsilon}^2/2(1+\epsilon)\right) = 0,
 \label{exact_anom2}
\end{gather}
where $D_\nu (z)$ is the parabolic cylinder function and $F(a, b, z)$ is the confluent hypergeometric function~\cite{NIST:DLMF}. 

\begin{figure}
\includegraphics[width=\columnwidth]{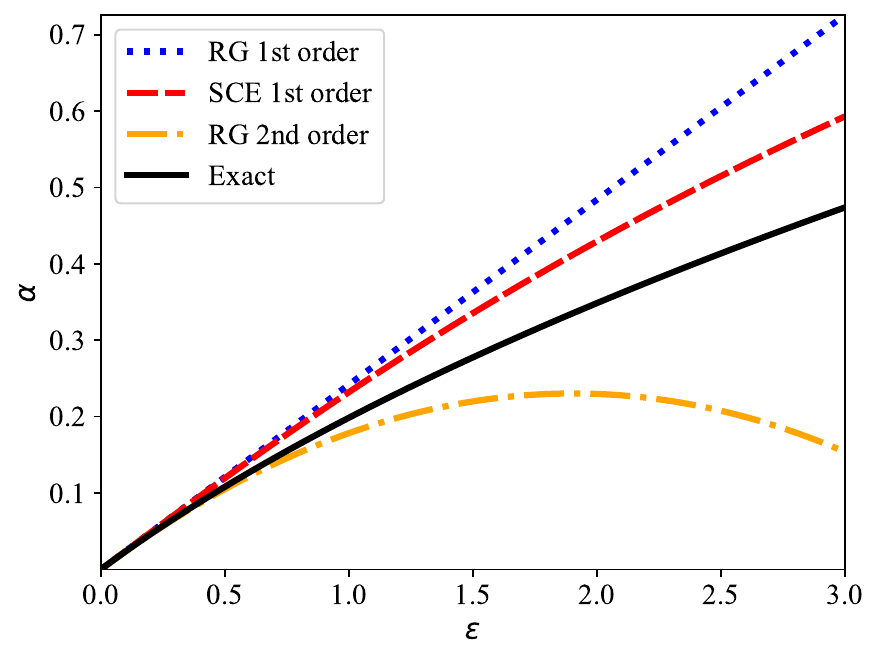} %revision4
\caption{\label{fig:comparison} Estimates of the anomalous dimension $\alpha$ as a function of the perturbation strength $\epsilon$ as calculated by renormalization group (first and second order in $\epsilon$), the self-consistent expansion at lowest order, and the exact value by solving Eq.~(\ref{exact_anom1}) and~(\ref{exact_anom2}) using a shooting method.} 
\end{figure}

Depicted in Fig.~\ref{fig:comparison}, we now compare the estimation of $\alpha(\epsilon)$ from 1st order SCE calculation~[Sec.~\ref{subsec:SCE_PDE}], 1st and 2nd RG calculations~\cite{goldenfeld1990anomalous,yoshida2005exact,cole1996self}, against the exact  results of Eq.~(\ref{exact_anom1}) and~(\ref{exact_anom2}).

In the regime of $\epsilon \ll 1$, all methods yield accurate estimations; for large $\epsilon$, the SCE method significantly improves the RG results, in particular avoiding the decrease in the anomalous dimension for $\epsilon > 2$ obtained by the second order in $\epsilon$ RG calculation. This finding again demonstrates the effectiveness of self-consistent methods for large perturbation problems, indicating their potential in strongly correlated physical systems.

\section{\label{sec:FT}RG and SCE in field-theoretic framework}
Recently, the SCE method had success in approximating Hamiltonian quantum mechanics and integral representations of special functions~\cite{remez2018divergent}. This problem, together with the single polymer chain executing a self-avoiding walk reviewed in Sec.~\ref{sec:Polymer}, is akin to a zero-dimensional field theory, and can be presented in an integral form for the partition function or generating function
\begin{align}
Z(x) \equiv \int ds \, e^{-H (x, s)} = \int ds \, e^{-H_0 (x, s) - \epsilon H_I (x,s)},
\end{align}
where $H_0$ is meant to represent the non-interacting problem, $H_I$ represents the interaction term. We will be expanding about the value of $s$ where $H_0$ has an extremum, representing a saddle point of the integral.  In the presence of $H_I$, the exact saddle point will of course move.  To implement the SCE approach, we will seek the best effective $H_0$ by introducing a new variational parameter $\eta$ to rescale $H_0$ at its maxima, obtaining a new perturbation theory
\begin{align}
Z(x) = \int ds \, e^{-H_\eta (x, s) - [-H_\eta (x, s) + H_0 (x, s) + \epsilon H_I (x,s)]},
\end{align}
where the square brackets will be denoted as $\delta H$, and are treated as a perturbation about the effective non-interacting Hamiltonian $H_\eta$. Next, we choose a self-consistent criterion, which states that the higher order correction of an observable of significance, which we will call $\hat g$ should vanish. For example, in the case of the single polymer reviewed earlier, the root mean square end-to-end distance or radius of gyration was chosen.  Then the expectation value of $\hat g$ will be 
\begin{align}
g(x) \equiv \langle \hat g(x,s)\rangle &= \frac{1}{Z}\int ds \, \hat{g} e^{-H (x, s)}\\
&\approx \frac{1}{Z_\eta}\int ds \, \hat{g}\, e^{-H_\eta (x, s)} \big[1-\delta H + O(\delta H^2) \big],
\end{align}
where the partition function $Z$ has been expanded as 
\begin{align}
    Z(x) &= \int ds \, e^{-H_\eta (x, s) - [-H_\eta (x, s) + H_0 (x, s) + \epsilon H_I (x,s)]}\\
    &= Z_\eta (1 + O(\delta H)).
    \label{eq:norm}
\end{align}

Note that the partition function $Z$ has also been expanded in powers of $\delta H$.
The choice of $H_\eta$ will be determined by the condition that in $g(x)$the terms of order $\delta H$ vanish.

All of this is straightforward conceptually when one is dealing with a partition function or generating function in a statistical field theory. But what happens when we are faced with a strong coupling dynamical system in the form of a PDE? This is the question addressed in the next section.

\subsection{From dynamical systems to field theory -- the MSR formalism}

For a dynamical systems problem, usually formulated in terms of differential equations, how can we find a systematic way to apply SCE? If the differential equations have a closed-form integral solution, the answer would be intuitive; but in most of cases, how to apply the SCE method, i.e. where to insert the variational parameter and how to choose an appropriate self-consistent criterion, is non-trivial. In order to make progress, we use the fact that it is possible to express the solution of a differential equation --- a deterministic function --- as the limit of a probability distribution.  Thus, our strategy is to convert the differential equation to a field theory using the so-called Martin-Siggia-Rose (MSR) formalism \cite{martin1973statistical, janssen1976lagrangean,bausch1976renormalized,tauber2014critical}, where the solution to differential equations becomes an expectation over a fluctuating field. For Barenblatt's equation, this strategy was first implemented by Yoshida \cite{yoshida2005exact}, who used the so-called exact or functional renormalization group to calculate the anomalous dimension to second order in $\epsilon$.

We introduce a random field $\phi (x, t)$ whose distribution peaks at the solution $u(x,t)$
\begin{align}\label{eq:random_field}
u(x, t) &= \int  D[\phi] \, \phi(x, t)\, \delta[u - \phi].
\end{align}
Then we rewrite the delta-functional constraint to ensure that the function $\phi (x,t)$ is a solution of Barenblatt's equation~(\ref{eq:barenblatt}) including the initial condition, and expand the delta-functional in the exponential representation, as a functional integral over an auxiliary field $\tilde{\phi}$, obtaining
\begin{widetext} 
\begin{align}\label{eq:MSR}
u(x, t) &= \int  D[\phi] \, \phi\, \delta\!\left[\partial_t\phi - \frac{\kappa}{2} \left[1 + \epsilon\,\Theta(-\partial_t \phi)\right] \partial_x^2 \phi - \delta (t) v_0(x)\right] \nn\\
        &= \int  D[\phi,\tilde{\phi}] \, \phi\, \exp\left\{- \int dy ds\, \tilde{\phi}\left[\partial_s - \frac{\kappa}{2} \left[1 + \epsilon\,\Theta(-\partial_s \phi)\right]\partial_y^2\right]\phi  + \int dy ds\,\delta (s) v_0(y)\tilde{\phi}\right\}.
\end{align}
The last line of Eq.~(\ref{eq:MSR}) suggests that we define the generating functional $Z[J, \tilde{J}]$ for the field $\phi$ and the associated action $S[\phi, \tilde{\phi}]$ as
\begin{align}
Z[J, \tilde{J}] &\equiv \int  D[\phi, \tilde{\phi}]\, e^{-  S[\phi, \tilde{\phi}]  + \int dy ds\,\left( J\tilde{\phi}+ \tilde{J}\phi \right) } \\
S[\phi, \tilde{\phi}] &\equiv\ - \int dy ds\, \tilde{\phi}\left[\partial_s - \frac{\kappa}{2} \left[1 + \epsilon\,\Theta(-\partial_s \phi)\right]\partial_y^2\right]\phi,
\end{align}
and thus  $u(x,t)$ can be expressed as the expectation value of the field
\begin{align}\label{eq:generate_u}
u(x, t) = \left\langle\phi (x, t)\right\rangle = \left. \frac{1}{Z} \frac{\delta Z}{\delta \tilde{J}} \right\rvert_{\tilde{J} = 0, J = \delta (t) v_0(x)}.
\end{align} 
This succinct reformulation of Barenblatt's equation within a field-theoretic framework allows us to proceed with standard field theory calculations, including RG and the SCE or both.

\subsection{Perturbation theory using generating functionals and Feynman diagrams}
Following Eq.~(\ref{eq:generate_u}), we first work out the usual perturbation theory in this framework using the generating functional method \cite{fradkin2021quantum}. For clarity, we write down the explicit expressions for the action and the generating functional:
\begin{align}
S[\phi, \tilde{\phi}] &\equiv S_0 [\phi, \tilde{\phi}] + S_\epsilon [\phi, \tilde{\phi}]
					 = \int dy ds\, \tilde{\phi}\left(\partial_s - \frac{\kappa}{2} \partial_y^2\right)\phi +  \int dy ds\, \tilde{\phi}\left[ - \frac{\epsilon\kappa}{2} \Theta(-\partial_s \phi)\partial_y^2\right]\phi \label{eq:PT_FT}\\
Z_0 [J, \tilde{J}] &\equiv \int  D[\phi, \tilde{\phi}]\, e^{-  S_0 + \int dy ds\,\left( J\tilde{\phi}+ \tilde{J}\phi \right) }
                   = A_0 \exp\left\{ \int dx dt\int dy ds \, \tilde{J}(x, t)\Delta_0(x,t; y,s) J(y,s)\right\}   \\
Z [J, \tilde{J}] &= \int  D[\phi, \tilde{\phi}]\, e^{-  S_0 - S_\epsilon  + \int dy ds\,\left( J\tilde{\phi}+ \tilde{J}\phi \right) }
                 = \exp\left\{-S_\epsilon \left[\frac{\delta}{\delta \tilde{J}}, \frac{\delta}{\delta J}\right]\right\} Z_0 [J, \tilde{J}].    
\end{align}
where $A_0$ is a constant factor and $\Delta_0(x, t; y, s)=e^{-\frac{(x-y)^2}{2 \kappa (t-s)}}/\sqrt{2\pi\kappa (t-s)}$ is the propagator for the zeroth order system. %revision6

At first order in $\epsilon$, we calculate the perturbation expansion for $Z [J, \tilde{J}]$ and evaluate at appropriate choice for the external fields $J$ and $\tilde{J}$
\begin{align}
Z [J, \tilde{J}] &= \left\{1 -S_\epsilon \left[\frac{\delta}{\delta \tilde{J}}, \frac{\delta}{\delta J}\right]\right\} Z_0 [J, \tilde{J}] \nn\\
    &= A_0 \left\{1 + \frac{\epsilon\kappa}{2}\int_0^t ds \int_{-X_0}^{X_0} dy\,  \frac{\delta}{\delta J(y, s)} \partial_y^2\frac{\delta}{\delta \tilde{J}(y, s)} \right\} \exp\left\{ \int dy_1 ds_1 dy_2 ds_2 \, \tilde{J}(y_1, s_1)\Delta_0(y_1,s_1; y_2,s_2) J(y_2,s_2)\right\} \nonumber \\
    &= A_0 \left\{1 + \frac{\epsilon\kappa}{2}\int_0^t ds \int_{-X_0}^{X_0} dy\, \left[\int d2 \, \partial^2_y \Delta_0(y, s; 2) J(2) \int d1 \, \tilde{J}(1) \Delta_0(1; y,s) + \left.\partial^2_y \Delta_0(y,s; y_2, s_2)\right\rvert_{y_2 = y, s_2=s} \right] \right\}e^{\int\tilde{J}\Delta_0 J} \nonumber \\
    &\xrightarrow[J = \delta (t) v_0(x)]{\tilde{J} = 0} A_0 \left\{1 + \frac{\epsilon\kappa}{2}\int_0^t ds \int_{-X_0}^{X_0} dy\,  \left.\partial^2_y \Delta_0(y,s; y_2, s_2)\right\rvert_{y_2 = y, s_2=s} \right\},
\end{align}
where we use ``$dn$'' as an abbreviation for ``$dy_n ds_n$.'' Similarly for the functional derivative
\begin{align}
\frac{\delta Z}{\delta\tilde{J} (x,t)}  &= Z\int d2 \, \Delta_0(x, t; 2) J(2) + A_0\frac{\epsilon\kappa}{2}\int_0^t ds \int_{-X_0}^{X_0} dy\int d2 \, \partial^2_y \Delta_0(y, s; 2) J(2) \Delta_0(x, t; y,s) e^{\int\tilde{J}\Delta_0 J} \nn \\
    &\xrightarrow[J = \delta (t) v_0(x)]{\tilde{J} = 0}  A_0 \left\{u_0 + \frac{\epsilon\kappa}{2} \int_0^t ds \int_{-X_0}^{X_0} dy\, \left[ u_0(x,t)\left.\partial^2_y \Delta_0(y,s; y_2, s_2)\right\rvert_{y_2 = y, s_2=s} +  \partial^2_y u_0(y,s) \Delta_0(x, t; y,s)\right] \right\}.
\end{align}
Utilizing cancellation of vacuum diagrams, we find the first order expansion of the required quantity $u(x,t)$ to be
%\begin{widetext}
\begin{align}\label{FT_PT}
u(x, t) &= \frac{u_0 + \frac{\epsilon\kappa}{2} \int_0^t ds \int_{-X_0}^{X_0} dy\, \left[ u_0(x,t)\left.\partial^2_y \Delta_0(y,s; y_2, s_2)\right\rvert_{y_2 = y, s_2=s} +  \partial^2_y u_0(y,s) \Delta_0(x, t; y,s)\right] +\mathcal{O}(\epsilon^2)}{1 + \frac{\epsilon}{2}\int_0^t ds \int_{-X_0}^{X_0} dy\,  \left.\partial^2_y \Delta_0(y,s; y_2, s_2)\right\rvert_{y_2 = y, s_2=s} + \mathcal{O}(\epsilon^2)} \nonumber\\
          & = u_0 + \frac{\epsilon\kappa}{2} \int_0^t ds \int_{-X_0}^{X_0} dy\,\partial^2_y u_0(y,s) \Delta_0(x, t; y,s)+\mathcal{O}(\epsilon^2) 
\end{align} 

Notice that we have recovered Eq.~(\ref{eq:naive_expansion}) in the PDE framework, with the same logarithmic divergence at the first order. Similarly, we can renormalize this divergence in the field theory framework.
\end{widetext}

\subsection{The Callan-Symanzik equation}
The initial width $l$ effectively serves as a cut-off, and our goal is to eliminate the divergence in $\left\langle\phi (x, t)\right\rangle$ in the limit $l \to 0$. Define $G^{(1)} (x, t; \epsilon, l) \equiv \left\langle\phi (x, t)\right\rangle$. We impose wave function renormalization with an arbitrary finite length scale $\mu$:
\begin{gather}
G^{(1)}_R (x, t; \epsilon, \mu)  = Z_{\phi}(\epsilon, \mu, l) \, G^{(1)} (x, t; \epsilon, l) \nonumber \\
\Rightarrow \, \phi_R = Z_{\phi} \, \phi, \,\tilde{\phi}_R = Z_{\phi}^{-1} \, \tilde{\phi}.
\end{gather}

The bare quantity $G^{(1)}$ is independent of the arbitrary length scale $\mu$, and therefore for any fixed $l$
\begin{align}
0 &= \mu \frac{d}{d\mu}  G^{(1)} (x, t; \epsilon, l)  \nn\\
&= \mu \frac{d}{d\mu} \left[Z_{\phi}^{-1}(\epsilon, \mu, l) \, G^{(1)}_R (x, t; \epsilon, \mu) \right],
\end{align}
which we solve to obtain the Callan-Symanzik equation
\begin{align} \label{FT_CS}
\left[-\frac{d \ln Z_{\phi}}{d \ln \mu} + \mu \frac{\partial}{\partial \mu}\right] G_R^{(1)} = 0
\end{align}
Again, we impose the renormalizability assumption 
\begin{gather}
 \lim_{l \to 0} \frac{d \ln Z_{\phi} }{d \ln \mu} = \text{dimensionless constant} \equiv 2 \alpha \nn\\
\Rightarrow \, Z_{\phi} = \left(\frac{\mu}{l}\right)^{2\alpha}.
\end{gather}
We note the steps outlined above closely parallel the renormalization procedure in Sec.~\ref{subsec:self_similar}, as we promised earlier. 

\subsection{Perturbative RG}
Recall that perturbative RG addresses divergences by resumming the perturbative solution, given a set of renormalization assumptions. Since both the perturbative solution~Eq.(\ref{FT_PT}) and the renormalization assumption~Eq.(\ref{FT_CS}) are identical to those discussed in the PDE framework in Sec.~\ref{subsec:PDE_RG}, the ensuing calculation is also exactly the same, and need not be repeated.

\subsection{SCE}
To apply SCE in this field-theoretic framework, we first construct the new perturbation theory starting with redefining the zeroth order system. When $\epsilon = 0$, we have
\begin{align}
Z_0 [J, \tilde{J} = 0] & = D[\phi, \tilde{\phi}]\, e^{-  S_0 + \int dy ds\,  J\tilde{\phi} }  \nonumber \\
& = \int  D[\phi, \tilde{\phi}]\, e^{-  \int dy ds\left(\tilde{\phi} \Delta_0^{-1}\phi - J\tilde{\phi} \right)} \\
u_0 (x,t) & = \int dy ds\, \Delta_0(x, t; y, s) J(y, s),
\end{align}
where $J(x, t) = \delta(t)\frac{Q_0}{\sqrt{2\pi l^2}} \exp\left(-\frac{x^2}{2 l^2}\right)$. 
Notice that we have two unperturbed terms to rescale, but they only appear in a multiplicative form in $u_0$ so we can choose one. Here we choose $J(x, t)$ to show the calculation.

We want the form of the self-consistent solution to be similar to the unperturbed case, so we replace the constant $Q_0$ with a free parameter $Q_\alpha$, and define $J_\alpha(x, t) = \delta(t)\frac{Q_\alpha}{\sqrt{2\pi l^2}} \exp\left(-\frac{x^2}{2 l^2}\right) $ such that when $\epsilon> 0$,
\begin{align}
u_\alpha (x,t) & = \int dy ds\, \Delta_0(x, t; y, s) J_\alpha(y, s) + \text{(h.o.t.)}
\end{align}
is an asymptotic solution. Therefore, the generating functional for the new zeroth order system is
\begin{widetext}
\begin{align}
Z_\alpha [J, \tilde{J}] &\equiv \int  D[\phi, \tilde{\phi}]\, e^{-  S_0 + \int dy ds\,\left( J_\alpha \tilde{\phi}+ \tilde{J}\phi \right) } 
        = A_0 \exp\left\{\int dx dt\int dy ds \, \tilde{J}(x, t)\Delta_0(x,t; y,s) J_\alpha(y,s)\right\}.
\end{align}
Using the rescaled generating functional, we construct the new perturbation theory
\begin{align}
Z [J, \tilde{J}] & = \int  D[\phi, \tilde{\phi}]\, e^{-  S_0 - S_\epsilon  + \int \left( J - J_\alpha\right)\tilde{\phi} +\int \left( J_\alpha\tilde{\phi}+ \tilde{J}\phi \right) } 
    = \exp\left\{-S_\epsilon \left[\frac{\delta}{\delta \tilde{J}}, \frac{\delta}{\delta J_\alpha}\right] + \int\left( J - J_\alpha\right)\frac{\delta}{\delta J_\alpha}\right\} Z_\alpha [J, \tilde{J}].
\end{align}
To first order, we calculate the new perturbation expansion
\begin{align}
Z [J, \tilde{J}] & = \left\{1 -S_\epsilon \left[\frac{\delta}{\delta \tilde{J}}, \frac{\delta}{\delta J_\alpha}\right] + \int\left( J - J_\alpha\right)\frac{\delta}{\delta J_\alpha} \right\} Z_\alpha [J, \tilde{J}] \nn\\
& = A_0 \left\{1 + \frac{\epsilon\kappa}{2}\int_0^t ds \int_{-X_\alpha}^{X_\alpha} dy\, \left[\int d2 \, \partial^2_y \Delta_0(y, s; 2) J_\alpha (2) \int d1 \, \tilde{J}(1) \Delta_0(1; y,s) + \left.\partial^2_y \Delta_0(y,s; y_2, s_2)\right\rvert_{y_2 = y, s_2=s} \right] \right\}e^{\int\tilde{J}\Delta_0 J_\alpha}  \nn\\
 &\xrightarrow[J_\alpha]{\tilde{J} = 0} A_0 \left\{1 + \frac{\epsilon\kappa}{2}\int_0^t ds \int_{-X_\alpha}^{X_\alpha} dy\,  \left.\partial^2_y \Delta_0(y,s; y_2, s_2)\right\rvert_{y_2 = y, s_2=s} \right\},
\end{align}
and similarly for the functional derivative 
\begin{align}
    \frac{\delta Z}{\delta\tilde{J} (x,t)}  
    &= Z\int d2 \, \Delta_0(x, t; 2) J_\alpha (2) 
    \!\begin{multlined}[t]
        + A_0\left\{ \frac{\epsilon\kappa}{2}\int_0^t ds \int_{-X_\alpha}^{X_\alpha} dy\int d2 \, \partial^2_y \Delta_0(y, s; 2) J_\alpha (2) \Delta_0 (x, t; y,s)  \right. \\
        \left. + \int dy ds \left[ J(y,s) - J_\alpha (y,s)\right]\Delta_0(x, t; y, s)\right\} e^{\int\tilde{J}\Delta_0 J} \nn
    \end{multlined}\\
    &\xrightarrow[J_\alpha]{\tilde{J} = 0}   A_0 \left\{u_\alpha 
         + \frac{\epsilon\kappa}{2} \int_0^t ds \int_{-X_\alpha}^{X_\alpha} dy\, \left[ u_\alpha (x,t)\left.\partial^2_y \Delta_0(y,s; y_2, s_2)\right\rvert_{y_2 = y, s_2=s} +  \partial^2_y u_0(y,s) \Delta_0(x, t; y,s)\right] + u_0 - u_\alpha \right\}.
\end{align}
After the cancellation of vacuum diagrams, we arrive at
%\begin{multline}
\begin{align}
u(x, t) = u_\alpha + \frac{\epsilon\kappa}{2} \int_0^t ds \int_{-X_\alpha}^{X_\alpha} dy\,  \partial^2_y u_\alpha (y,s) \Delta_0(x, t; y,s) 
+ u_0 - u_\alpha + \text{(h.o.t.)},
\end{align}
%\end{multline}
where the first order correction matches its counterpart in the PDE framework~(\ref{SCE_u1}) precisely
\begin{align}
u_\alpha^{(1)} = -u_\alpha + u_0 
+ \frac{\epsilon}{2} \int_0^t ds\int_{-X_\alpha (s)}^{X_\alpha (s)} dy\,\Delta_0 (x, y; t, s) \,\partial_y^2 u_\alpha(y,s). 
\end{align}
Subsequent application of the self-consistent criterion follows the established procedure in Sec.~\ref{subsec:SCE_PDE}, and therefore does not need to be repeated here.  In other words, the field-theoretic formalism conveniently allows the RG and SCE methods to be systematically carried out to an arbitrary order of perturbation theory.
\end{widetext}

\section{\label{sec:conclusion}Conclusion and future work}
In this paper, we have extended the SCE method to study a well-known dynamical problem, Barenblatt's equation for underground water spreading in a porous medium. We presented an analytical SCE calculation in two frameworks, a PDE framework and a field-theoretic framework, establishing their equivalence. 
Our calculation has two steps.  The first is to use general renormalization group considerations to obtain a Callan-Symanzik equation that describes how anomalous dimensions may arise in the equation.  The second step is to use an explicit SCE perturbation calculation constrained by the Callan-Symanzik equation to derive the anomalous dimension.  Remarkably, even a rapid calculation at  first order allows our SCE method to provide an accurate estimation of the anomalous dimension $\alpha (\epsilon)$. Notably, in the regime of large perturbations (large $\epsilon$), our results improve upon the existing RG results. We were not able to obtain these results using either the RG or the SCE alone.

The original application of RG to Barenblatt's equation presaged its eventual use for moving boundary problems with self-similarity of the second kind in turbulence propagation \cite{chen1992renormalization}, traveling waves \cite{chen1994renormalization,chen1995numerical,paquette1994structural} and the full spectrum of singular perturbation problems \cite{chen1994global,chen1996renormalization,veysey2007simple}.  Although the RG approximants for singular perturbation problems are remarkably accurate even when the supposedly small parameter $\epsilon = O(1)$, there is still a need to be able to generate approximants at even larger values.  For example, in low Reynolds number fluid dynamics, the problem of flow around a body leads to extremely challenging asymptotics problems with boundary layers that even RG approximations \cite{veysey2007simple} have difficulty in extending beyond Reynolds numbers of order unity.  Thus, it would be of great interest to develop improved singular perturbation techniques for all these problems based on SCE. Additionally, adapting SCE for numerical schemes is a compelling direction, given the difficulties singular problems pose for analytical approaches.

The potential scope of the SCE method extends beyond physics. As more successful application examples emerge in various problems, its rigorous mathematical foundations and convergence theory are ripe for further exploration~\cite{remez2018divergent}. In computational science, particularly within physics-informed machine learning, the principle of self-consistency is becoming increasingly relevant. Lin \el introduced a deep learning solver to solve the polymer self-consistent field theory equations~\cite{lin2022deep}. Shen \el combined the self-consistency in the Fokker-Planck equation with neural networks to achieve convergent solutions and facilitate efficient stochastic gradient descent~\cite{shen2022self}, and the approach was further generalized to solve other PDEs~\cite{li2024self}.  It would be interesting to leverage the SCE method within machine learning to tackle complex and nonlinear physics dynamics, focusing on efficient computation, accuracy, and robust convergence, even in the presence of singular perturbations.

%\nocite{*}
\medskip

\begin{acknowledgements}
We thank Elie Raphael for correspondence regarding the use of SCE in polymer problems. This work was partially supported by a grant from the Simons Foundation (Grant number 662985, NG).
\end{acknowledgements}

\bibliography{SCE}% Produces the bibliography via BibTeX.

\end{document}